\documentclass[twocolumn,preprintnumbers,amsmath,amssymb]{revtex4}
\usepackage{amsfonts}% Physical Review B
\usepackage{epsf}
\usepackage{anysize}
\marginsize{0.8 in}{0.8 in}{1in}{1.4in}
\usepackage{float}
\usepackage{graphicx}% Include figure files
\usepackage{dcolumn}% Align table columns on decimal point
\usepackage{bm}% bold math
\usepackage{amsmath}
\usepackage{amssymb}
\usepackage{mathrsfs}
\usepackage{pstricks,pst-node,pst-text,pst-3d}
\usepackage{subfigure}
\usepackage[all]{xy}
\usepackage{color}
\usepackage{epsf}
\usepackage{anysize}
\usepackage{graphicx}% Include figure files
\usepackage{dcolumn}% Align table columns on decimal point
\usepackage{bm}% bold math
\usepackage{amsmath}
\usepackage{amssymb}
\usepackage{mathrsfs}
\usepackage{pstricks,pst-node,pst-text,pst-3d}
\usepackage[all]{xy}
\usepackage{mathtools}
\usepackage{appendix}

\DeclarePairedDelimiterX\braket[2]{\langle}{\rangle}{#1 \delimsize\vert #2}

\makeatletter
  \newcommand\figcaption{\def\@captype{figure}\caption}
  \newcommand\tabcaption{\def\@captype{table}\caption}
\makeatother

\newcommand{\dd}{{\rm d}}

\begin{document}

%\preprint{APS/123-QED}

\title{Learning to Calibrate Quantum Control Pulses by Iterative Deconvolution}

\author{Xi Cao,$^{1,\dagger}$ Bing Chu,$^2$ Haijin Ding,$^1$ Luyan Sun,$^3$ Yu-xi Liu$^{4,5}$ and Rebing Wu$^{1,4,}$}
\thanks{
	rbwu@tsinghua.edu.cn\\$^\dagger$zuozhu24@outlook.com}
\address{ \small$^{1}$Department of Automation, Tsinghua University, Beijing 100084, China\\
	\small$^{2}$School of Electronic and Computer Science, University of Southampton, Southampton SO17 1BJ, United Kingdom\\
	\small$^{3}$Center for Quantum Information, Institute for Interdisciplinary Information Sciences, Tsinghua University, Beijing 100084, China\\
	\small$^{4}$Center for Quantum Information Science and Technology, BNRist, Beijing 100084, China\\
	\small$^{5}$Institute of Micro-Nano Electronics, Tsinghua University, Beijing 100084, China}

%\address{Department of Automation, Tsinghua University, Beijing, 100084, China}

\begin{abstract}
	%% Text of abstract
In experimental control of quantum systems, the precision is often hindered by imperfect applied electronics that distort control pulses delivered to target quantum devices. To mitigate such error, the deconvolution method is commonly used for compensating the distortion via an identified convolutional model. However, its effectiveness is limited by model inaccuracies (e.g., imprecise parameters or unmodeled distortion dynamics). In this paper, we propose a learning-based scheme to eliminate the residual calibration error by repeatedly applying the deconvolution operations. The resulting iterative deconvolution method is shown to be able to correct both linear and nonlinear model errors to the highest precision allowed by available finite sampling rates. The calibration error induced by finite sampling rates is also analyzed, from which we propose that the inter-sampling error can be suppressed by actively introducing nonlinear components in the control electronics.
\end{abstract}
\maketitle

%\tableofcontents
%

\section{Introduction}
\begin{figure}[]
	\centering
	\includegraphics[width=1\columnwidth]{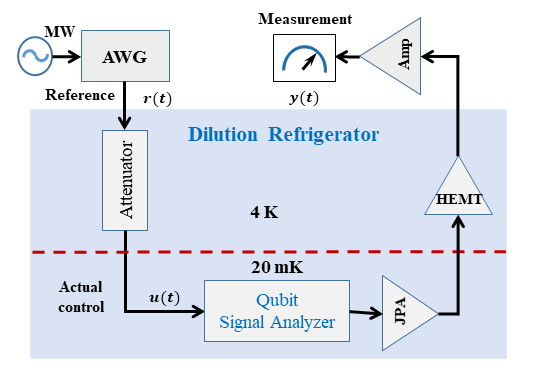}
	\caption{The transmission process of input signals. The reference signal $r(t)$ generated by AWG is distorted when being delivered through the transmission line to the target placed in the refrigerator. The {\it in situ} signal $u(t)$ is readout via a qubit from $y(t)$.}
	\label{system}
\end{figure}

Towards practical applications of quantum information processing technology \cite{nielsen2010quantum}, high precision control of quantum state and gate operations is the core enabling technology \cite{dong2010quantum,1367-2630-12-7-075008}. To date, high-fidelity quantum gates above error-correction threshold have been achieved \cite{PhysRevLett.117.060504}, but there is still a long way to go for scalable quantum computation due to the decoherence induced by environmental interactions and systematic errors induced by imperfect control electronics. In this paper, we are concerned with the latter systematic error caused by the distortion of control signals delivered to the qubits \cite{PhysRevApplied.4.024012,PhysRevA.84.022307,SPINDLER201249,Glaser2015,PhysRevX.8.011030,PATTERSON2006231,article}. For example, in the manipulation of superconducting qubits shown in Fig.~\ref{system}, the arbitrary waveform generator (AWG) sends control signals to the superconductor quantum circuits via transmission lines \cite{123}, which, together with the associated transmission line and electronic components, can induce linear distortion by spurious inductance, capacitors and resonators, as well as nonlinear distortion induced by current-dependent inductances and resistances \cite{PhysRevApplied.4.024012,PhysRevA.84.022307,SPINDLER201249,Glaser2015,PhysRevX.8.011030,doi:10.1063/1.4866691,stephen}. Besides, in low-temperature experiments, the change of electronic properties of the circuit elements may lead to additional distortions \cite{PATTERSON2006231,article}. All these factors can severely decrease the precision of states and gate manipulations and hence must be corrected.

To compensate the pulse distortion, one can incorporate the error model into the optimization process for control pulses design \cite{PhysRevApplied.4.024012}. Alternatively, one can also directly calibrate the pulse to a designed shape, e.g., the deconvolution method that has been applied to the flux bias control of superconducting quantum circuits \cite{johnson2011controlling,jansson2014deconvolution,olkin2009edited}. In both approaches, an identified model is required for quantifying the pulse distortion, and its accuracy determines how much the distortion can be compensated. 

Under the circumstance that an accurate model is not always available, the pulse calibration must learn from the measured error signals instead of merely using the model. In quantum domain, the earliest application was in the iterative learning control of chemical reactions with ultrafast laser pulses \cite{PhysRevLett.68.1500}, and later was extended to atomic and optical systems, and further to information processing \cite{Long2014,0953-4075-44-15-154013,PhysRevLett.112.240503,R1,PhysRevX.8.031086}. These studies were in principle ``black-box" learning without using any {\it a priori} information (i.e., the model) about the quantum control system, which is different with classical ``grey-box" iteration learning control (ILC) that learns more efficiently by incorporating a (even coarse) model. The latter has been widely implemented in classical control engineering \cite{bristow2006survey,owens2005iterative}, such as industrial robots \cite{4048052,999653}, computer numerical control machine tools \cite{485814} and autonomous vehicles \cite{doi}, and recently to quantum control for online tuneup of high-precision quantum gates \cite{PhysRevA.97.042122}.

In this paper, we introduce the model based ILC to the calibration of {\it in situ} signals based on the standard (offline) deconvolution. The remainder of this paper is organized as follows. Section \ref{Sec:II} will introduce the deconvolution method, following which the iterative deconvolution method is proposed and demonstrated by numerical simulations. In Section \ref{Sec:III}, we analyze the influence of the inter-sampling oscillation induced by finite sampling rate, and the stability of the learning process. In Section \ref{Sec:IV}, we show that the nonlinearity in the distortion can be actively used for suppressing the inter-sampling oscillation. Finally, Section \ref{Sec:V} draws the conclusion.

\section{Iterative Deconvolution: Methods and Simulations}\label{Sec:II}

In this section, we will introduce the deconvolution method and show how it can be improved by iterative learning.

\subsection{Deconvolution}

Suppose that we wish to find a proper AWG signal $r(t)$ that produces a desired {\it in situ} control signal $u_d(t)$ to the quantum system, which cannot be achieved by directly setting $r(t)=u_d(t)$, because the yielded output pulse $u(t)$ will be distorted by the control transmission line that goes from room temperature to the low temperature.

The idea of deconvolution is to identify a linear convolutional input-output model:
\begin{equation}\label{}
u(t) = \int_0^\infty \bar{g}(t-\tau)r(\tau)\dd \tau
\end{equation}
for the distortion of $r(t)$, where $\bar{g}(t)$ is the impulse response of the distortion. In Laplace domain, the convolutional model can be described by transfer function as:
\begin{equation}\label{}
u(s) = \bar{G}(s)r(s),
\end{equation}
where $r(s)$, $u(s)$ and $\bar{G}(s)$ are the Laplace transform of $r(t)$, $u(t)$ and $\bar{g}(t)$. When the actual dynamics of the distortion is also linear, say $G(s)$, then by setting the reference signal $r(s) = \bar{{G}}^{-1}(s)u_d(s)$, the produced {\it in situ} control signal is $u(s)=G(s)\bar{G}^{-1}(s)u_d(s)$. Apparently, the desired signal can be perfectly produced, only when the identified model is precise, i.e., $G(s)=\bar{G}(s)$. This inverse-system based method for compensating convolutional distortion is called deconvolution.

\subsection{From deconvolution to iterative deconvolution}
In practice, the precision of deconvolution calibration is always limited due to imprecise identified parameters or unmodeled linear or nonlinear dynamics in $\bar{G}(s)$. In the following, we will show that the resulting residue signal errors can be corrected by repeatedly using the identified imprecise model and online observation of the error signal.

In the following, we simply assume that the {\it in situ} signal $u(t)$ has been precisely perceived. Note that this is non-trivial, because $u(t)$ needs to be reconstructed from some qubit readout signal $y(t)$ (e.g., by Ramsey experiments \cite{hofheinz2008generation}), but we will not delve into this issue.
Denote the initial AWG signal by $r^{(0)}(t)$, and the distorted input signal is thus $u^{(0)}(s)=G(s)r^{(0)}(s)$. According to the error $e^{(0)}(t) = u_d(t)-u^{(0)}(t)$, we can modify the AWG signal using the error signal and the reference model $\bar{G}(s)$, as follows
\begin{equation}
r^{(1)}(s)=r^{(0)}(s) + \beta \bar{G}^{-1}(s)e^{(0)}(s),
\label{inverse}
\end{equation}
where $\beta$ is the learning rate that needs to be sufficiently small for the stability of the iteration. The updated AWG signal is tested by the system, following which the error can be obtained for the next iteration of calibration. Inductively, we can repeat this process by updating the AWG signal with the error signal until the iteration converges. The iterative application of deconvolution compensation will be called iterative deconvolution.

According to Eq.~(\ref{inverse}), it is easy to derive that:
\begin{equation}
e^{(k+1)}(s)=\left[I-\beta G(s)  \bar{G}^{-1}(s)\right]e^{(k)}(s).
\label{IMID}
\end{equation}
Therefore, if one can manage to keep the operator norm $\|I-\beta G(s)  \bar{G}^{-1}(s)\|$ smaller than 1, the iterative deconvolution method will guide the AWG input to:
\begin{equation}\label{eq:final r(t)}
  r(s)=G^{-1}(s)u_d(s)
\end{equation}
that perfectly yields {\it in situ} signal $u(t)=u_d(t)$. The convergence holds when $\beta$ is sufficiently small and when the model error is not large (i.e., $G(s)\bar{G}^{-1}(s)$ is reasonably close to identity) \cite{harte2005discrete}. Moreover, as long as the iteration converges, the chosen reference model only affects the rate of convergence, but not on final yield (\ref{eq:final r(t)}). Therefore, the iterative deconvolution is by nature immune to the model imprecision.

\subsection{Simulation Results}
 In the following simulations, we choose the step function as the desired signal to be produced {\it in situ}, which is required for fast quantum switching operation \cite{PhysRevB.78.104508}.

\begin{figure}[]
	\centering
	\includegraphics[width=1\columnwidth]{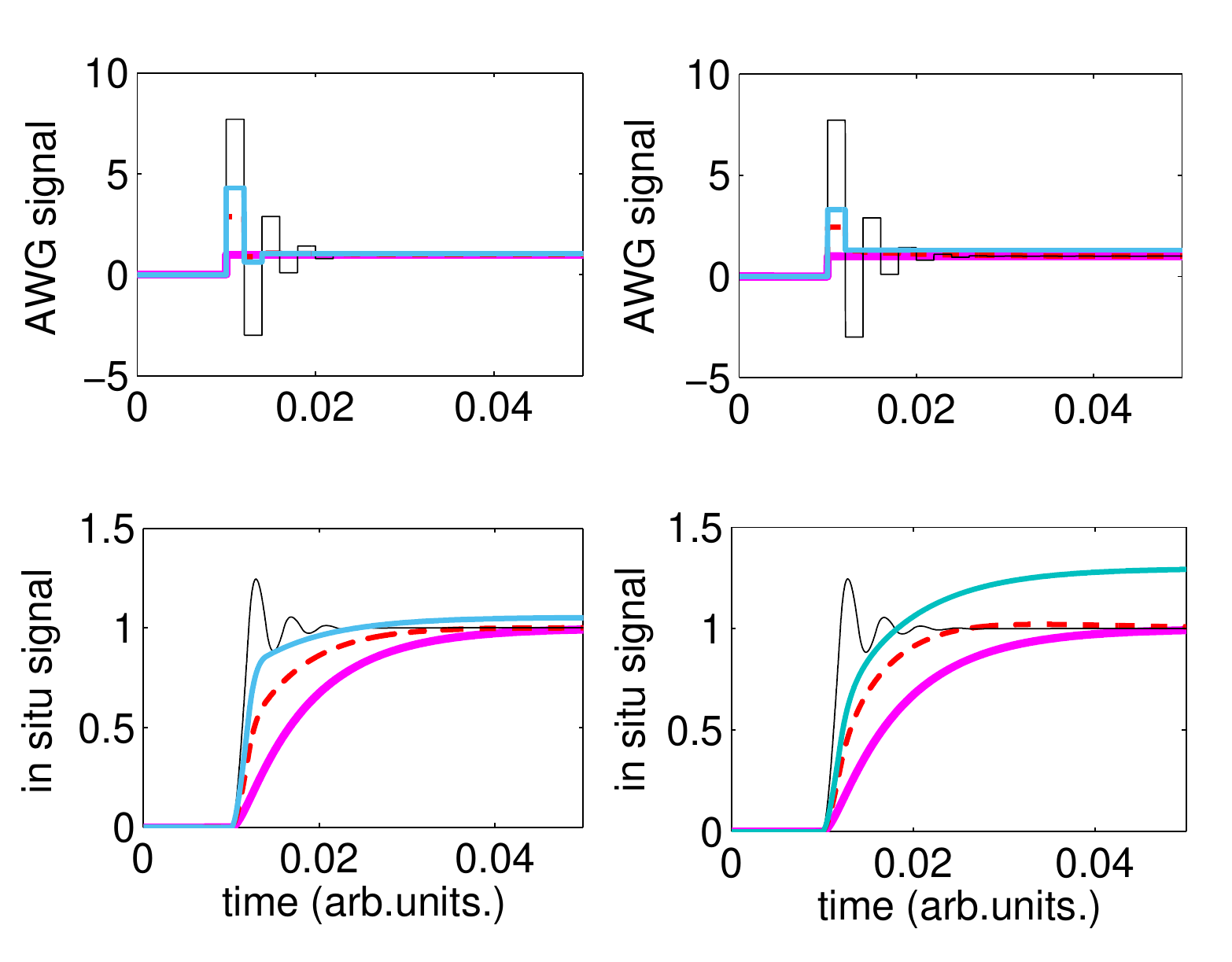}
	\caption{The AWG signals and {\it in situ} signals using the good model $\bar{G}_1(s)$ (left plots) and the bad model $\bar{G}_2(s)$ (right plots), where the AWG sampling period is $\tau = 0.002$ (arb.units.). The AWG signal is initially chosen as a step function (purple) and the calibrated signals (after 100 iterations) are shown in black. The dash red curves correspond to the intermediate results (in the 2nd iteration) and the blue curves are obtained by using the standard non-iterative deconvolution.}
	\label{small_large_2sa} %% label for entire figure
\end{figure}

We start from the simpler case in which the real system is described by the following linear transfer function:
\begin{equation}
{G}(s)=\frac{1}{(0.008s+1)(0.001s+1)},
\label{system_model}
\end{equation}
which involves a slow part (characterized by $T_1=0.008$) and a fast part (characterized by $T_2=0.001$). These parameters are chosen only for illustration, and they vary under different physical circumstances. We start the test with a ``good" reference model:
\begin{equation}
\bar{{G}}_1(s)=\frac{1}{(0.006s+1)(0.001s+1)},
\label{model1}
\end{equation}
in which only $T_1$ is slightly different, and a ``bad" reference model:
\begin{equation}
\bar{{G}}_2(s)=\frac{1}{0.004s+1},
\label{model2}
\end{equation}
in which not only $T_1$ is very imprecise, but also the fast dynamics is ignored. We also take into account the discretization effect of the AWG device, i.e., the implementable $r(t)$ signals are always piecewise constants with a sampling rate $\tau$. For example, we pick the sampling period $\tau=0.002$ (arb.units.) and simulate the iterative deconvolution for 100 iterations with learning rate $\beta = 0.5$, as shown in  Fig.~\ref{small_large_2sa}. It can be clearly seen that the iterative deconvolution takes the {\it in situ} signal from a slowly rising shape to a fastly rising shape that is much closer to the the desired step function, no matter which reference model is used, and the same reference input $r(t)$ is obtained. The only difference between using good and bad models is the shape of signals during the intermediate iterations (see the dash red curves). The iteration converges more slowly when using the bad model, but not so much, than that based on the good model. Figure \ref{small_large_2sa} also compares the calibration results with non-iterative deconvolution, under which the performance is much worse, because the performance of non-iterative calibration is heavily dependent on the accuracy of the reference model.

\begin{figure}[]
	\centering
		\includegraphics[width=\columnwidth]{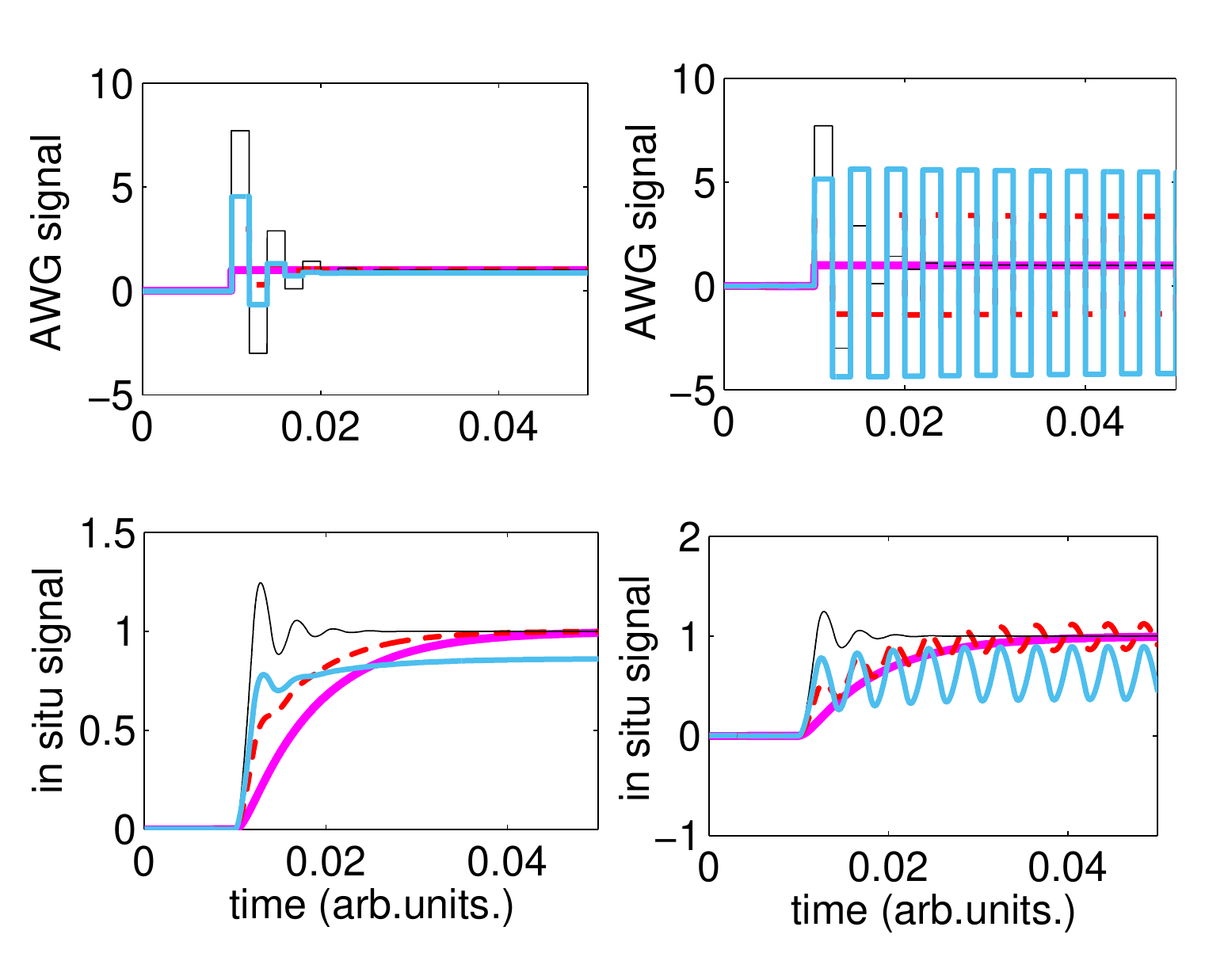}
	\caption{The AWG signals and {\it in situ} signals using non-minimum-phase models $\bar{G}_3(s)$ (left plots) and $\bar{G}_4(s)$ (right plots), where the AWG sampling period is $\tau = 0.002$ (arb.units.). The AWG signal is initially set as a step function (purple) and the calibrated signals (after 100 iterations) are shown in black. The dash red curves correspond to the intermediate results (in the 2nd iteration) and the blue curves are obtained by only using deconvolution.}
	\label{non_min_2sa} %% label for entire figure
\end{figure}

It should be noted that iterative deconvolution may fail when the model is too bad, in which case the iteration diverges. In particular, the iteration is more likely instable when the distortion dynamics is non-minimum phase (i.e., the transfer function contains zeros {or poles} with positive real parts \cite{franklin1994feedback}). In Fig.~\ref{non_min_2sa}, we simulate the calibration process using the following two non-minimum phase reference models:
\begin{eqnarray}
\bar{G}_3(s)&=&\frac{-0.002s+1}{(0.006s+1)(0.001s+1)},
\label{non1}\\
\bar{G}_4(s)&=&\frac{-0.006s+1}{(0.006s+1)(0.001s+1)}
\label{non2}
\end{eqnarray}
that contain zeros in the right half of complex plane. They yield the same AWG input as obtained with $\bar{G}_1(s)$ and $\bar{G}_2(s)$. However, the iterated signal exhibits strong oscillations with $\bar{G}_4(s)$ for many iterations, which shows that the calibration process is almost instable. The iterative can completely lose stability when the zero is closer to the imaginary axis, which is not shown here.

The above observation shows that one must be cautious when using a non-minimum phase model. Under such circumstance, the above inverse-system based iterative deconvolution (IMID) may never succeed, and one can turn to more stable norm-optimal iterative learning control, which is essentially a combination of inverse model algorithm and gradient-based algorithm. Interested readers are referred to \cite{amann1996iterative,owens2005iterative} for more details.

\section{Error analysis}\label{Sec:III}
The above simulations show that the iterative deconvolution can well outperform the deconvolution itself. However, the eventual calibrated pulse is never, more or less, precisely identical to the desired step function. Typically, overshoots and damping oscillations appear at the beginning of the calibrated {\it in situ} signal, due to the finite AWG sampling rate. In this section, we will analyse the origin and influences of such inter-sampling oscillations, as well as the stability of the iterative learning process.
\subsection{The inter-sampling oscillation}
\begin{figure}[]
	\centering
	\includegraphics[width=1\columnwidth]{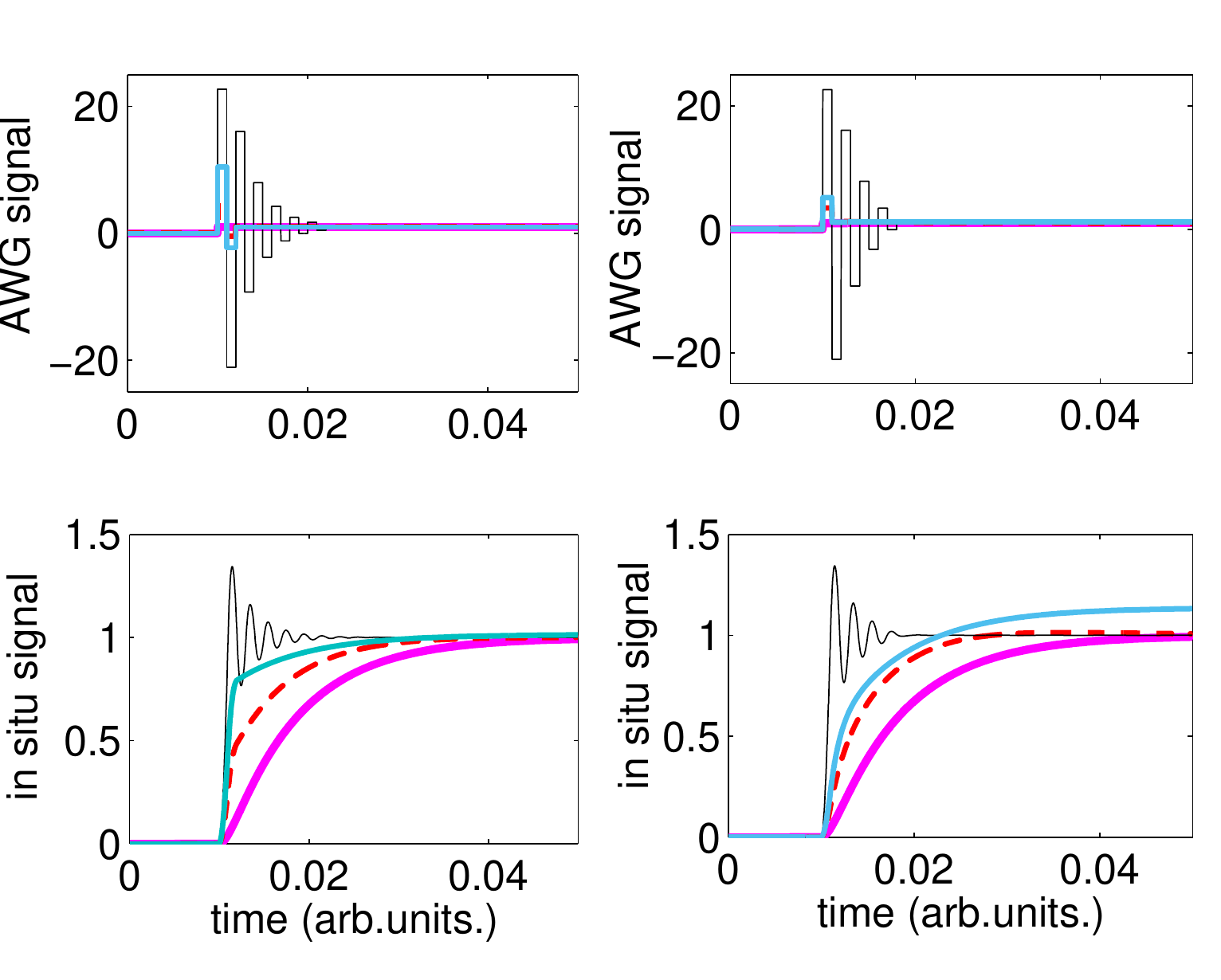}
	\caption{The AWG signals and {\it in situ} signals using $\bar{G}_1(s)$ and $\bar{G}_2(s)$, where the AWG sampling period is $\tau = 0.001$ (arb.units.). The AWG signal is initially set as a step function (purple) and the final curves (black) after 100 iterations are shown in black. The dash red curves correspond to the intermediate results (in the 2nd iteration) and the blue curves are obtained by non-iterative deconvolution. }
	\label{small_large_1sa} %% label for entire figure
\end{figure}

Before analyzing the origin of the observed inter-sampling oscillations, let us examine how they vary under a faster sampling period $\tau = 0.001$ (arb.units.). As shown in Fig.~\ref{small_large_1sa}, the iteration also successfully converges, but with higher overshoot and longer oscillations. 

It is not hard to understand how this happens. In the first sampling period, the driving field $r(t)$ attempts to drive the {\it in situ} signal $u(t)$ from 0 to 1 at the first sampling time point $t=\tau$, which requires a high power when the sampling period is short. Due to the inertial dynamics of $G(s)$, $u(t)$ will keep going up after crossing 1 at $t=\tau$. Then, the driving field $r(t)$ switches to pull $u(t)$ back to 1 at the second sampling time point $t=2\tau$, and again the inertia effect brings $u(t)$ down below 1 after $t=2\tau$. On and on, the inter-sampling oscillations persist and gradually damp. Only when $G(s)$ is first-order, there is no inter-sampling oscillation (see Appendix). 

\begin{figure}[]
				\centering
				\includegraphics[width=\columnwidth]{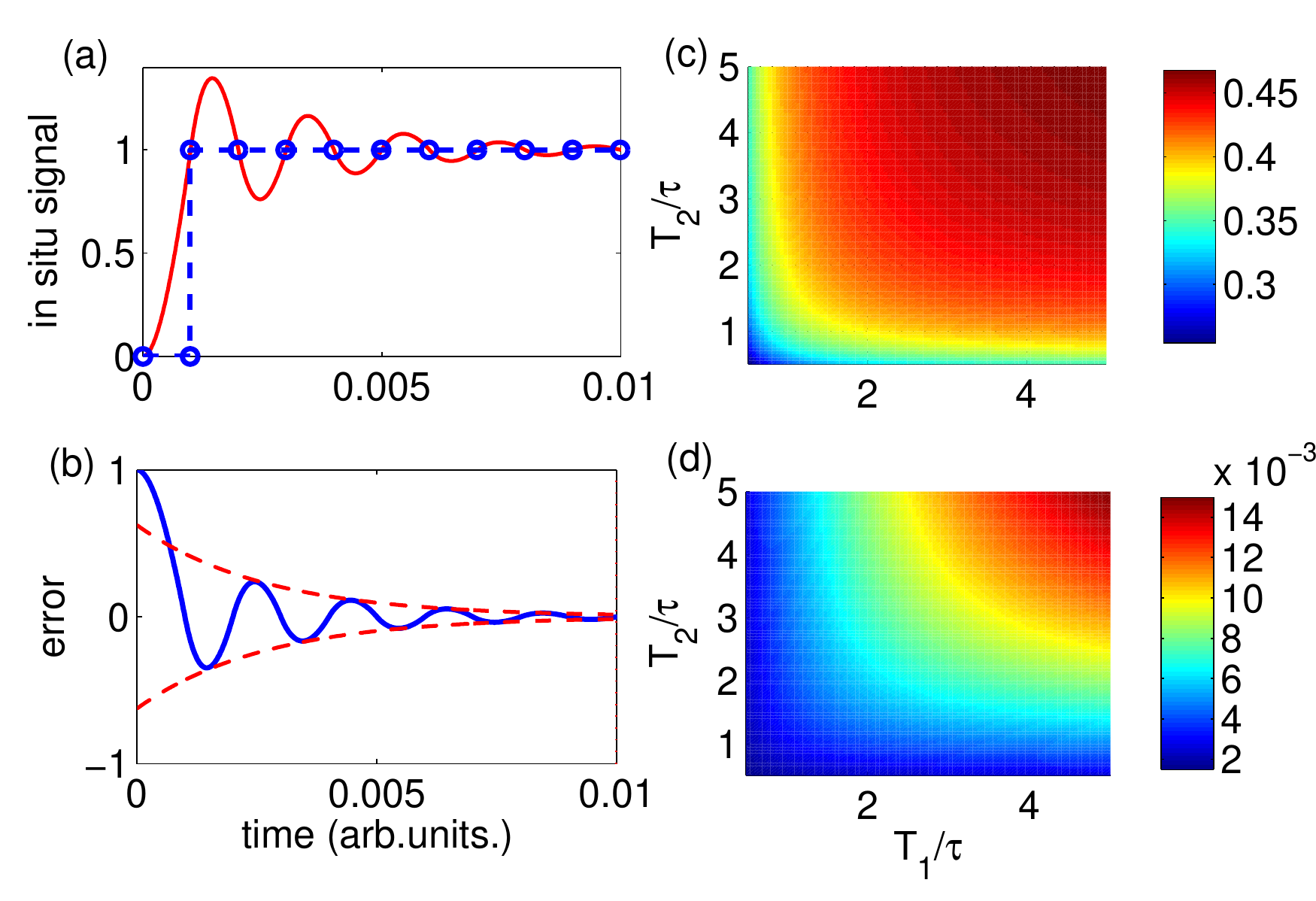}
				\caption{(a).~The calibrated {\it in situ} signal under sampling rate $\tau=0.001$ (arb.units.). The signal matches the reference signal perfectly at the sampling points, but is deviated from it in between; (b).~The blue curve corresponds to the error between continuous {\it in situ} signal and the desired in (a), the dashed red lines are the fitted exponential decaying curves; (c).~the decay time constant of the inter-sampling oscillation; (d).~the overshoot  of the inter-sampling oscillation.}
				\label{sampled_analysis} %% label for entire figure
			\end{figure}

The inter-sampling behavior is essentially determined by the distortion dynamics $G(s)$, which is independent with the reference model used for calibration. Taking second-order $G(s)$ for examples, we show in Figs.~\ref{sampled_analysis}(c) and \ref{sampled_analysis}(d) the dependence of the highest overshoot and the damping time ($T_s$ in Eq.~(\ref{B})) on the time constants $T_1$ and $T_2$ (defined in Eq.~(\ref{second})) of $G(s)$. The inter-sampling oscillation is severer when $T_1$ and $T_2$ are larger, implying that the distortion dynamics should be as fast as possible in order to mitigate the inter-sampling errors.

\begin{figure}[]
	\centering
	\includegraphics[width=1\columnwidth]{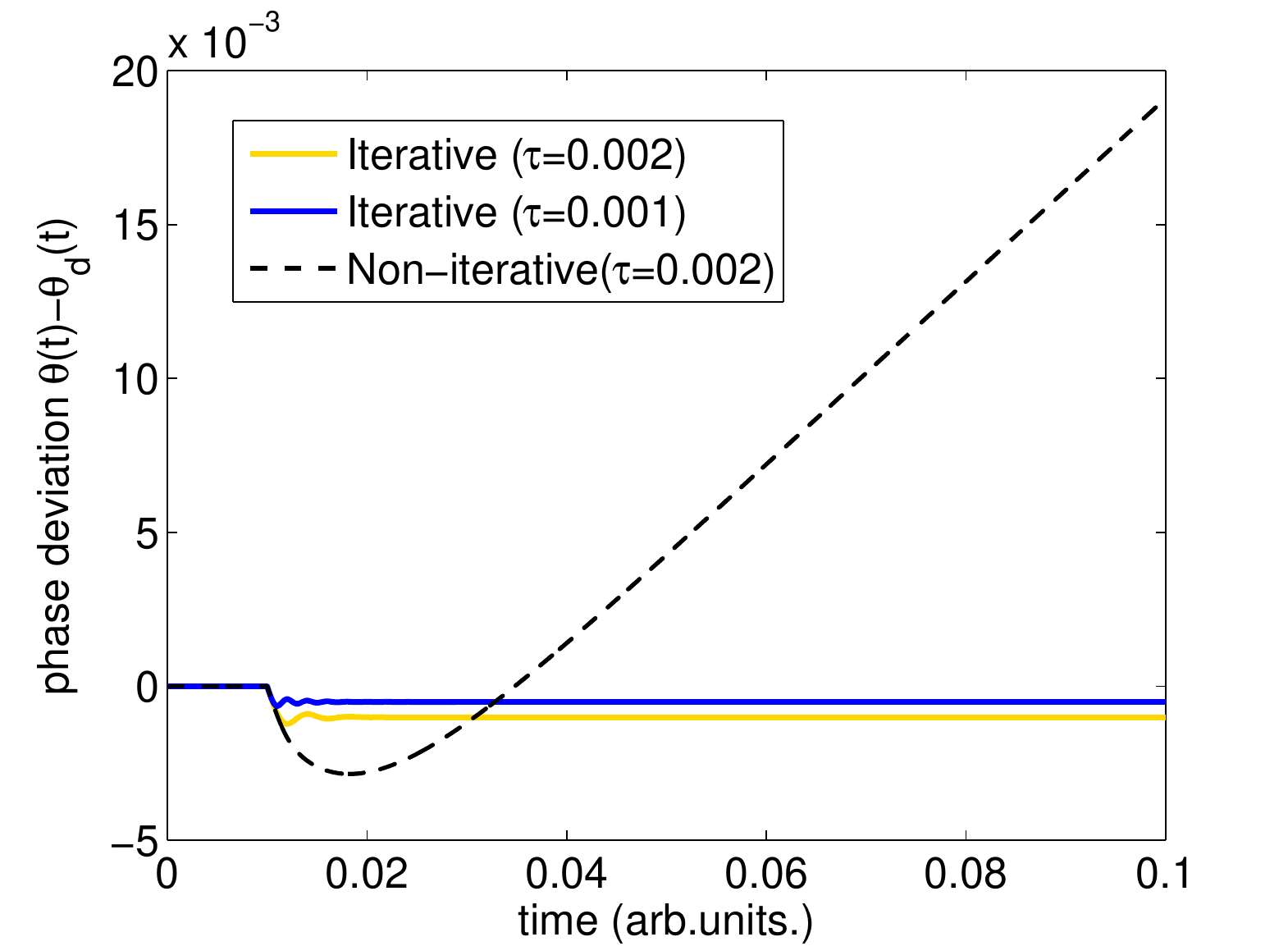}
	\caption{The accumulated phase deviation of the qubit probe from that under ideal flux bias tuning, where the reference model $\bar{G_2}(s)$ is used for iterative and non-iterative deconvolution calibration.}
	\label{quantum_state} %% label for entire figure
\end{figure}

To see how the calibrated AWG signal affect the measured qubit readout signal that is used to reconstruct the {\it in situ} signal $u(t)$, we take the example of flux bias tuning (i.e., $u(t)$ ) on a flux superconducting qubit \cite{johnson2011controlling}, in which the phase of the qubit is readout through a Ramsey experiment \cite{PhysRev.78.695,doi:10.1002/prop.200310063}. Mathematically, the readout signal $y(t) = \cos \theta(t)$, where the accumulated qubit phase is the integral of the {\it in situ} signal $u(t)$, as follows:
\begin{equation}\label{}
 \theta(t)= \int_0^t u(\tau){\rm d}\tau.
\end{equation}
Ideally, $\theta(t)$ should follow the linear rising function $\theta_d(t)=t$.
For simplicity, we restrict our discussion within a time interval that $\theta(t)\in[0,\pi]$, so that $u(t)$ can be uniquely determined. Figure \ref{quantum_state} shows that the phase deviation from $\theta_d(t)$ can be kept very small when using iterative deconvolution calibration, which is much better than the non-iterative calibration. Due to the inter-sampling effect, the steady-state error in $\theta(t)$ is nonzero and it decreases when using higher sampling rates. Moreover, it can be seen that the inter-sampling oscillation in the phase $\theta(t)$ is much weaker after being averaged out by integration.

\subsection{Sampling-time and continuous-time errors}
\begin{figure}[]
				\centering
				\includegraphics[width=1\columnwidth]{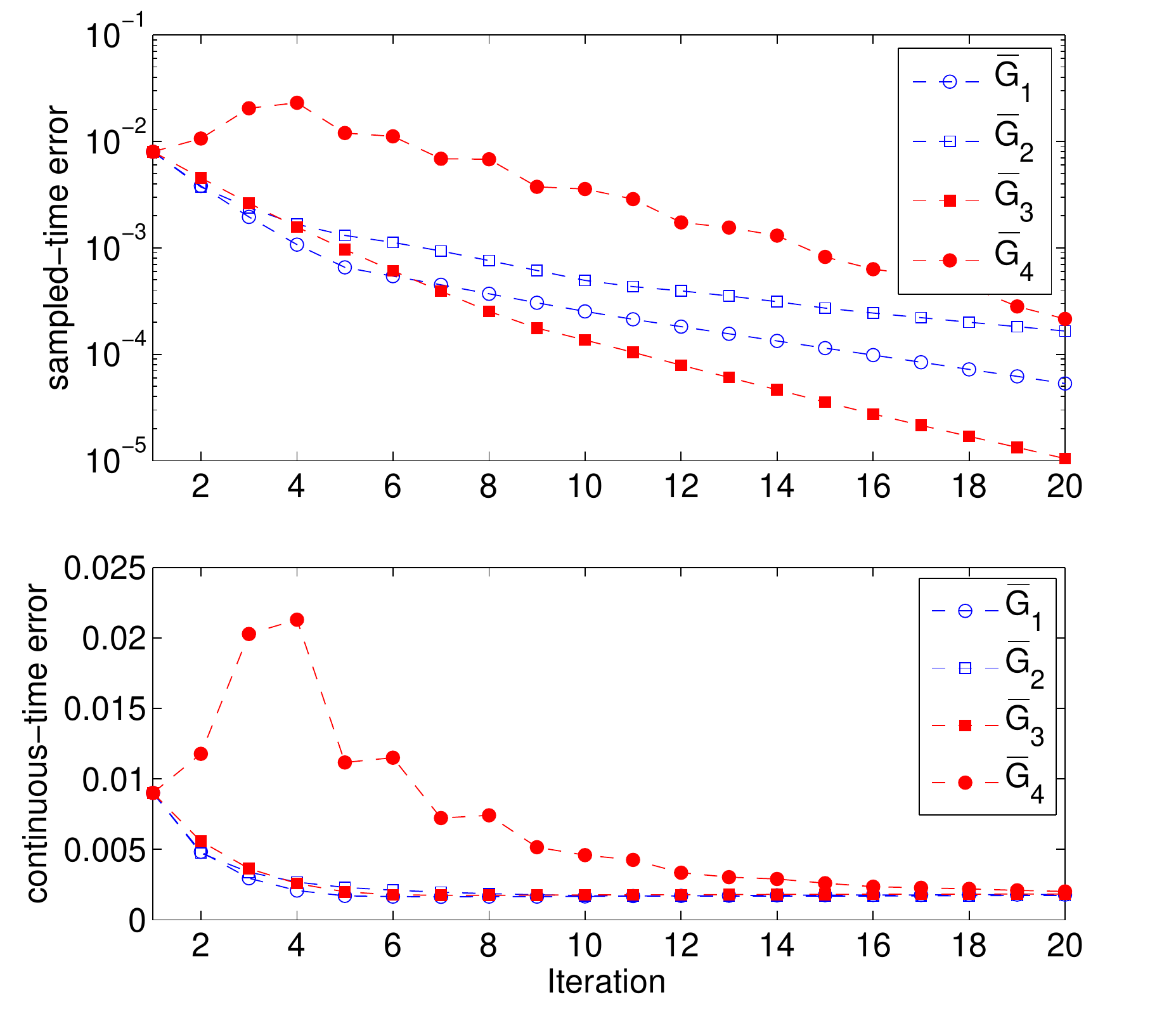}
				\caption{(a).~The sampled errors for the iterative deconvolution calibration using models $\bar{G}_1(s)$, $\bar{G}_2(s)$, $\bar{G}_3(s)$ and $\bar{G}_4(s)$; (b).~The continuous-time errors for the iterative deconvolution calibration using models $\bar{G}_1(s)$, $\bar{G}_2(s)$, $\bar{G}_3(s)$ and $\bar{G}_4(s)$.}
				\label{error}
			\end{figure}

In Fig.~\ref{sampled_analysis}(a), we find that the calibrated {\it in situ} signal after applying iterative deconvolution perfectly matches $u_d(t)$ at every sampling point except the first one. This makes sense because we can only correct what we can see. Therefore, the physically existing overshoots and oscillations is invisible to the sampled-data measurements, which are assumed to be done at the same rate of AWG. In this regard, we define the sampling-time and continuous time errors as follows:
\begin{eqnarray}
\mathcal{E}_{\rm sample}&=&\sum^{N}_{k=1}[u(k\tau)-u_d(k\tau)]\tau.
\label{error_definition}\\
\mathcal{E}_{\rm continuous}&=&\int_0^T|u(t)-u_d(t)|\dd t,
\label{error_definition1} 
\end{eqnarray} 
where $T$ is the duration of the input signal and $N$ is the number of sampling points.
Figure~\ref{error} shows how these two types of errors vary in the iteration processes. The sampled error can, as expected, be made arbitrarily small, but the actual errors converge to finite values due to the inter-sampling oscillation. Besides, under the same learning rate ($\beta = 0.5$), the convergence is faster when using more precise models. When the model is slightly non-minimum phase, the convergence can be accelerated a little, but this advantage does not hold when the iteration process is close to instability (e.g., when using $\bar{G}_4(s)$ for calibration). 

\subsection{The stability of iterative learning}

\begin{figure}[]
	\centering
	\subfigure{
		\includegraphics[width=1\columnwidth]{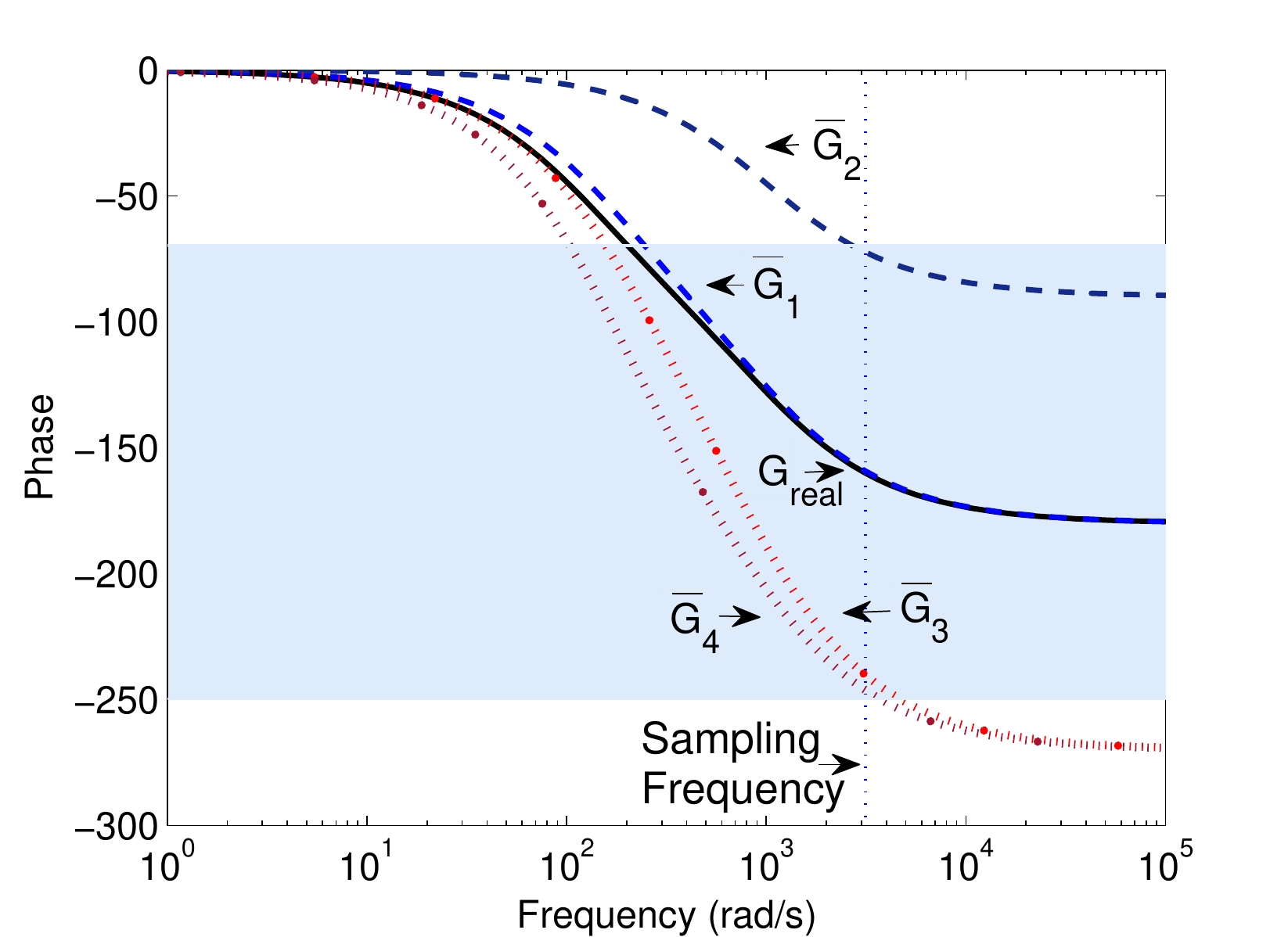}}
	\caption{The phase plots for the actual system and the model $\bar{G}_1(s)\sim\bar{G}_4(s)$ used for iterative deconvolution. The non-minimum-phase models ($\bar G_3(s)$ and $\bar G_4(s)$) converges relatively faster but may lose stability because they are closer to the border of the stability region (shaded area).}
		\label{bode} %% label for entire figure
\end{figure}

We have shown that the iterative deconvolution works when the reference error model is not accurate. However, the error model cannot be too inaccurate, otherwise the iterative learning will diverge. Figure~\ref{bode} explains how the stability of the iterative learning relies on the model from its phase-frequency property. As indicated in \cite{harte2005discrete}, the iteration is stable when the phase difference between the reference model and the real model is within 90 degrees at the sampling frequency (see the shaded area). The convergence becomes faster when the phase of the real model $G(i\omega)$ is in advance of that of the reference model $\bar{G}(i\omega)$ (e.g., see Fig.~\ref{error} for $\bar G_3(s)$ and $\bar G_4(s)$). However, the iterative deconvolution starts to oscillate when the phase of $\bar{G}(i\omega)$ is close to the border of stable region. Furthermore, when the phase is out of the region, the iterative learning will become unstable, which is very likely for non-minimum-phase models.

\section{Iterative Deconvolution in presence of nonlinearity}\label{Sec:IV}
In this section, we study how the iterative deconvolution works when the distortion of the signal is nonlinear.

\begin{figure}[]
	\centering
	\includegraphics[width=\columnwidth]{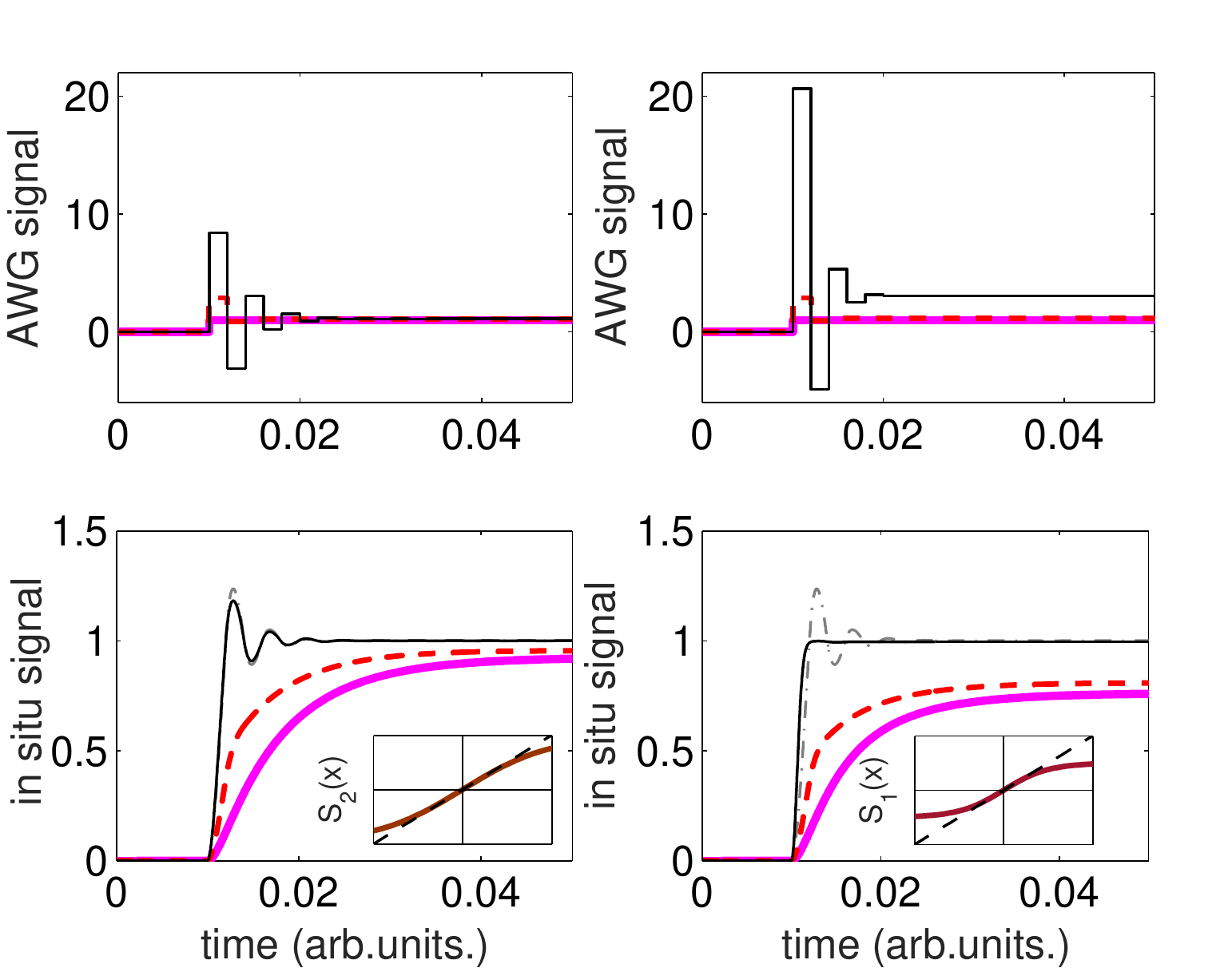}
	\caption{The AWG signals and {\it in situ} signals using small and large saturation for the model $\bar{G}_1(s)$, where the AWG sampling period is $\tau = 0.002$ (arb.units.). The AWG signal is initially set as a step function (purple) and the final curves without saturation (dash black) and with saturation (black) after 100 iterations are shown in black. The dash red curves correspond to the intermediate results (in the 2nd iteration).}
	\label{saturation_2sa} %% label for entire figure
\end{figure}

For demonstration, we assume that, after a linear distortion $G(s)$, the signal also experiences the following saturation nonlinearity:
\begin{equation}\label{}
  S_A(x) = A\tanh\left(\frac{x}{A}\right),
\end{equation}
where $A$ is the saturation bound. In the simulations shown by Fig.~\ref{saturation_2sa}, the saturation bound is chosen as $A=2$ and $A=1$, respectively, and we perform the same iterative algorithm using the linear reference model $\bar{G}_1(s)$ for deconvolution.  In both cases, the iteration deconvolution can still correct the error very well without including the nonlinearity in the reference model. When the saturation is relatively small (e.g., when $A=2$ shown in the figure), and the final calibrated {\it in situ} signal is only slightly different from the case without saturation. However, when the saturation is properly chosen (i.e., $A=1$), the calibrated {\it in situ} signal is remarkably different, but in a good way that the inter-sampling overshoots and oscillations are almost completely suppressed. The resulting calibration performance is much better than the linear case. Such side-effect implies that one can actively introduce nonlinearity into the distortion dynamics  to improve the calibration performance limited by finite sampling rates.

\begin{figure}[H]
	\centering
		\includegraphics[width=1\columnwidth]{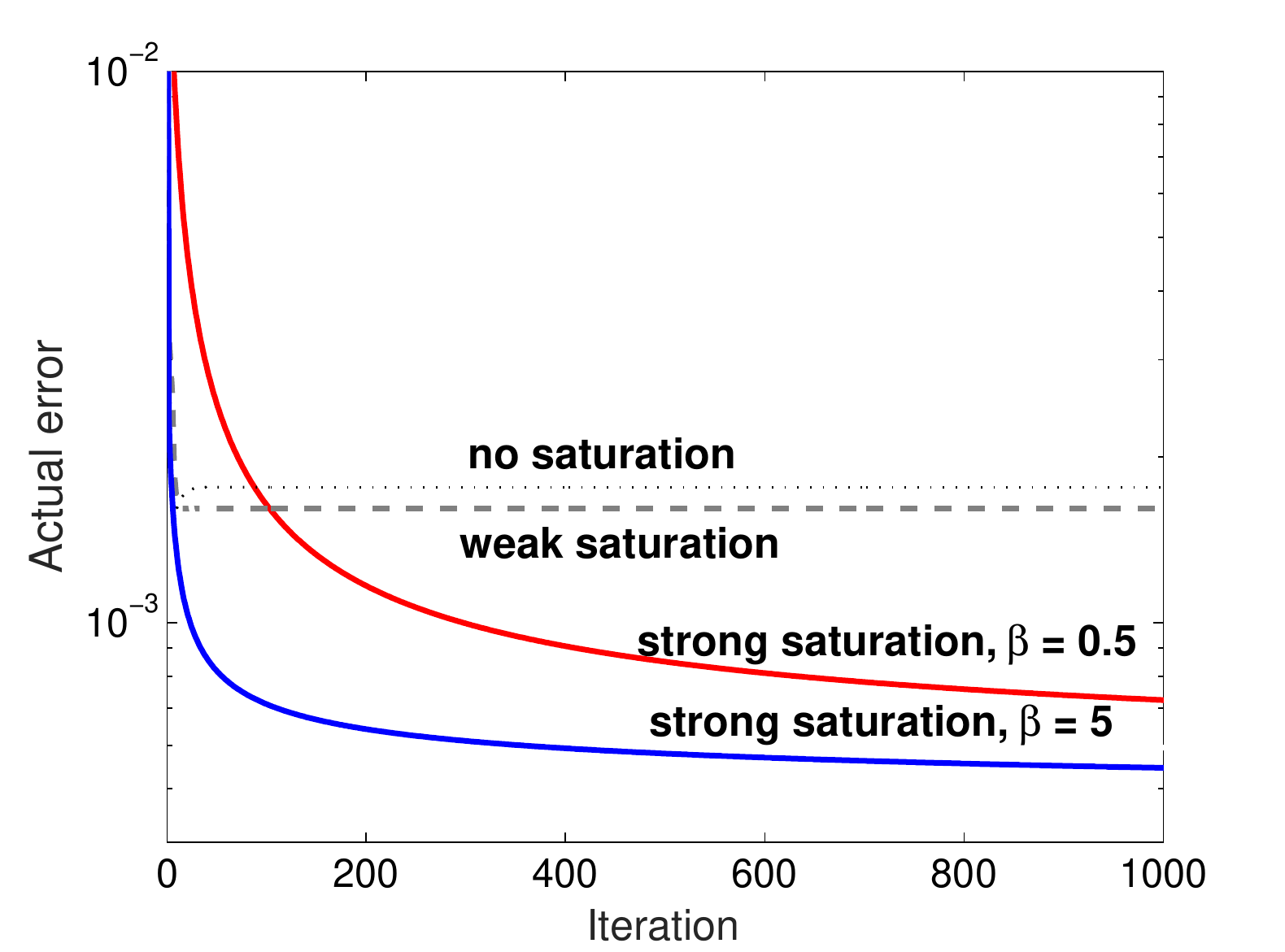}
	\caption{The calibration errors in the iteration process. The active use of saturation nonlinearity can further reduce the calibration error, and one can accelerate the learning process by tuning up the learning rate under which the iteration is still stable.}
	\label{SA_error} %% label for entire figure
\end{figure}

The saturator introduced above also makes the learning process much more stable, by which one can choose much larger learning rate to accelerate the learning process. As shown in Fig.~\ref{SA_error}, the learning under the rate $\beta=0.5$ takes about 300 iterations to reduce the error down to $\mathcal{E}_{\rm contiunous} = 10^{-3}$. For comparison, by increasing the learning rate to $\beta=5$, the learning process is still stable and achieve the same precision within just 20 iterations, resulting in a much faster convergence.

\section{Conclusion}\label{Sec:V}
To conclude, we propose an iterative learning deconvolution method for calibrating quantum control pulses. The simulation results demonstrate that such iterative deconvolution can effectively mitigate the residue error of deconvolution brought by model inaccuracies. The algorithm is robust and the ultimate calibration performance is limited by inter-sampling oscillations induced by the finite sampling rate. Such oscillation may be effectively suppressed by active use of nonlinearity in the electronic circuits.

Our work demonstrates the power of ``grey-box" learning that incorporates an imperfect model that is thought to be useless in traditional ``black-box" learning. Such learning process can be much more efficient using {\it  a priori} knowledge from the model that is even not so good. Note that in most of quantum information processing systems, an imperfect but still good model is usually not hard to construct, and one should active exploit the model to improve the precision and efficiency of the control design. This methodology can be extended to reduce the number of very costly experiments in many other quantum control problems (e.g., quantum gate tune-up \cite{PhysRevA.97.042122}).
\section{ACKNOWLEDGMENTS}
The authors acknowledge support from national Key Research and Development Program of China (Grant No. 2017YFA0304300) and NSFC grants (Nos. 61833010 and 61773232).

	\addcontentsline{toc}{chapter}{\appendixname}
	\begin{appendix}
		\section{The analysis of inter-sample behavior}
		For illustration, we assume that the real system is modeled by
		$$G(s) = \frac{(\tau_1s+1)(\tau_2s+1)\cdots(\tau_ms+1)}{(T_1s+1)(T_2s+1)\cdots(T_ns+1)},$$
		where $T_1>T_2>\cdots>T_n$ and the time constants in the transfer function are not precisely known.
		
		Suppose that the required AWG signal is:
		\begin{equation}
		\begin{aligned}
		r(t)&=R_11(t)+\sum_{k=1}^\infty (R_{k+1}-R_k)1(t-k\tau),
		\end{aligned}
		\end{equation}
		where $\tau$ is sampling period and $R_n$ is the magnitude of the signal during the $n$-th sampling period. Let $h(t)$ be the step+ response of $G(s)$, then the {\it in situ} signal $u(t)$ in the first $n$ sampling period can be derived as:
		\begin{equation}
		\begin{aligned}
		u(t,R_1,...,R_n)&=R_1h(t)+\sum_{k=1}^\infty (R_{k+1}-R_k)h(t-k\tau).
		\end{aligned}
		\label{ap1}
		\end{equation}
		For desired signal $u_d(t)$, (\ref{ap1}) poses $n$ linear equations of $R_1$, $R_2$, ... and $R_n$ as follows:
		\begin{equation}
		U(k\tau,R_1,R_2)=u_d(2\tau),\ \ \ k=1,2, ... ,n
		\label{A1}
		\end{equation}
		
		When $u_d(t)=u(t)$, it can be easily proven that for first order system $R_1=1/h(\tau)$ and $R_1=R_2=R_3...=R_n$, and the derivative of $u(t)$ after the first sampling point is 0. It implies iterative deconvolution can perfectly correct errors with first-order plants after the first sampling period.
		
		For second order systems:
		\begin{equation}
		G(s)=\frac{1}{(T_1s+1)(T_2s+1)},
		\label{second}
		\end{equation}
		we use the function:
		\begin{equation}
		f(t)=1+Ae^{-t/T_s}\sin(\frac{2\pi}{T}t),
		\label{B}
		\end{equation}
		to fit the oscillations of second order systems, in which $T_s$ corresponds to the damping time of the inter-sampling oscillations and $A$ corresponds to the magnitude of overshoot. Figs.~5(b) and 5(c) depict the dependence of the settling time and the overshoot on $T_1$ and $T_2$ when $\tau=0.001$ (arb.units.).
	\end{appendix}
\
\
\
\
\
\
\\
\\
\\
\\
\\
\\
\\
\\
\\
\\
\\
\\
\\
\\
\\
\\
\\
\\
\\
\\
\\
\\
\\
\\
\\

\bibliography{sample1}

\end{document}